\documentclass[galaxies,article,accept,moreauthors,pdftex,10pt,a4paper]{mdpi}
\usepackage{upgreek}

\firstpage{1}
\articlenumber{7}
\issuenum{1}
\pubvolume{7}
\pubyear{2019}
\copyrightyear{2018}
\history{Received: 7 November 2018; Accepted: 25 December 2018; Published: 29 December 2018}
\updates{yes}

\Title{Verification of Photometric Parallaxes with Gaia DR2~Data}

\Author{
Oleg Malkov $^{1,}$*\orcidA{},  
Sergey Karpov $^{2,3,4}$\orcidB{},
Dana Kovaleva $^{1}$\orcidC{},
Sergey Sichevsky $^{1}$\orcidD{},
\mbox{Dmitry Chulkov $^{1}$\orcidE{}},
Olga~Dluzhnevskaya $^{1}$\orcidF{},
Alexey Kniazev $^{3,5,6,7}$\orcidG{},
Areg Mickaelian $^{8}$\orcidH{},
Alexey~Mironov $^{7}$,
Jayant~Murthy $^{9}$\orcidJ{},
Alexey~Sytov $^{1}$\orcidK{},
Gang Zhao $^{10}$\orcidL{}
\mbox{and
Aleksandr Zhukov $^{1,7,11}$}
}

\AuthorNames{
Oleg Malkov,
Sergey Karpov,
Dana Kovaleva,
Sergey Sichevsky,
Dmitry Chulkov,
Olga~Dluzhnevskaya,
Alexey Kniazev,
Areg Mickaelian,
Alexey~Mironov,
Jayant~Murthy,
Alexey~Sytov,
Gang Zhao
and
Aleksandr Zhukov}

\address{%
$^{1}$ \quad Institute of Astronomy of RAS, 119017 Moscow, Russia; dkovaleva@mail.ru (D.K.); s.sichevskij@gmail.com~(S.S.); chulkov@inasan.ru (D.C.); olgad@inasan.ru (O.D.); sytov@inasan.ru (A.S.); aozhukov@mail.ru (A.Z.) \\  
$^{2}$ \quad Institute of Physics, Czech Academy of Sciences, 182 21 Prague, Czech Republic; karpov.sv@gmail.com \\ 
$^{3}$ \quad Special Astrophysical Observatory of RAS, 36916 Nizhnij Arkhyz, Russia; akniazev@saao.ac.za \\ 
$^{4}$ \quad Laboratory ``Fast Variable Processes in the Universe'', Kazan Federal University, 420008 Kazan, Russia \\
$^{5}$ \quad South African Astronomical Observatory, P.O. Box 9, Cape Town 7935, South Africa\\
$^{6}$ \quad Southern African Large Telescope Foundation, P.O. Box 9, Cape Town 7935, South Africa\\
$^{7}$ \quad Sternberg Astronomical Institute, 119234 Moscow, Russia; almir@sai.msu.ru\\
$^{8}$ \quad NAS RA V. Ambartsumian Byurakan Astrophysical Observatory, Byurakan 0213, Republic of Armenia; aregmick@yahoo.com\\
$^{9}$ \quad Indian Institute of Astrophysics, Bengaluru  560 034, India; jmurthy@yahoo.com\\
$^{10}$~~~~Key Laboratory of Optical Astronomy, National Astronomical Observatories, Chinese Academy of Sciences, Beijing 100012, China; gzhao@nao.cas.cn\\
$^{11}$~~~~Russian Technological University (MIREA), 119454 Moscow, Russia}
\corres{Correspondence: malkov@inasan.ru; Tel.: +7-495-951-7993}

\abstract{Results of comparison of {\it Gaia} DR2 parallaxes with
data derived from a~combined analysis of
2MASS (Two Micron All-Sky Survey),
SDSS (Sloan Digital Sky Survey),
GALEX (Galaxy Evolution Explorer), and
UKIDSS (UKIRT Infrared Deep Sky Survey)
surveys in four selected high-latitude $|b|>48^{\circ}$ sky areas are presented.
It is shown that
multicolor photometric data from large modern surveys can be used for
parameterization of stars closer than 4400 pc
and brighter than $g_{SDSS} = 19.^m6$,
including estimation of parallax and interstellar extinction value.
However, the stellar luminosity class should be properly determined.}


\begin{document}


\def\apj{Astrophys.~J}
\def\aaps{Astron.~and Astrophys.~Suppl.~Ser}
\def\pasp{Publ.~Astron.~Soc.~Pac}
\def\aap{Astron.~Astrophys}
\def\apjl{Astrophys.~J.~Lett}
\def\apjs{Astrophys.~J.~Suppl.~Ser}
\def\aj{Astron.~J}
\def\mnras{Mon.~Not.~R.~Astron.~Soc}
\def\apss{Astrophys.~Space.~Sci}

\section{Introduction}

\textls[-10]{One of the main problems of astrophysics is the study of the physical
properties belonging to the surface layers of stars. Because these stars are observed through
interstellar dust, their light is dimmed and reddened, complicating
their parameterization and classification. The parameters of a~given star
(temperature, gravity, metallicity, etc.), as well as the interstellar reddening, may be
obtained from its optical, infrared, and ultraviolet spectrum. However, one
must either use large telescope or bright objects to get spectral energy
distributions with good resolution and sufficient accuracy.
For~instance, spectroscopic observations with 1-h exposure time on a~2-m telescope
with low ($R \sim 1000$) and high ($R \sim$ 100,000) resolution
allow limiting magnitudes of 16--17 mag and 11--12~mag, respectively. An~8-m
telescope adds 3 mag to these estimations.
This~work was performed by various authors, and a~number
of empirical atlases were constructed
{
(\mbox{\citet{1972VilOB..35....3S}},
\citet{1992A&AS...92....1G},
\mbox{\citet{1996BaltA...5..603A}},
\citet{1997BaltA...6..481A},
\citet{1998PASP..110..863P},
\citet{2003Msngr.114...10B},
\mbox{\citet{2003A&A...402..433L}},
\mbox{\citet{2004ApJS..152..251V},}
\citet{2007ASPC..374..409H},
\citet{2011A&A...532A..95F},
\mbox{\citet{2011A&A...525A..71W}).}
}
However \mbox{\citet{2014BaltA..23..286M}} made a~critical analysis, compared data
for stars included in several atlases, and found
many discrepancies.}

Another way to construct a~map of interstellar extinction is its
estimation (as well as stellar parameters)
from evolutionary tracks. Corresponding procedures were developed in \citet{2016BaltA..25...67S}, \citet{2016ARep...60..816S}, \citet{2017ARep...61..193S} and applied to LAMOST data by \citet{2017AstBu..72...51S}.
However, a~knowledge of stellar atmospheric parameters is highly
desirable for the application of these procedures, limiting the number of stars available for such a~parameterization.

Therefore, the solution of the problem of parameterization of
stars based on their photometry is a~topical issue~\cite{2003BaltA..12..514M}.
A great variety of photometric systems (see, e.g.,
\citet{1992msp..book.....S}
for references) and
recently constructed large photometric surveys
(like SDSS~\cite{2011ApJS..193...29A} and GALEX~\cite{2005ApJ...619L...1M})
as well as VO-tools for
cross-matching surveys' objects provide with a~possibility to get
multicolor photometric data for millions of objects. Consequently, it
allows the user to parameterize objects and determine interstellar
extinction in the galaxy.

A comparative analysis of available 3D maps of interstellar
extinction was made by \citet{1997ARep...41...10K},
and contradictory results were found.
As a~temporary solution of that problem,
a~synthetic map of the galactic interstellar extinction can be compiled (see, e.g.,
\citep{1997BaltA...6..358K,1997AJ....114.2043H,2002Ap&SS.280..115M}).

Early dust maps used the correlation between the dust column
density and the distribution of neutral hydrogen \citep{1982AJ.....87.1165B}.
These data were supplanted by the dust maps produced by \mbox{\citet{1998ApJ...500..525S}}, who used full
sky microwave data made available by the IRAS (Infrared Astronomical Satellite)
mission and the
DIRBE (Diffuse Infrared Background Experiment) instrument on
the COBE (Cosmic Background Explorer) mission. Mapping the dust column densities via the calibrated
dust temperature, the extinction maps, assuming a~standard
reddening law, were shown to be at least twice as accurate as those
of \citet{1982AJ.....87.1165B}.
An advantage of this method is that it
does not rely on a~predefined model for the stellar population.

The successful implementation of the European astrometric space mission {\it Gaia}
(the second version of the mission catalog, {\it Gaia} DR2, was published in
April 2018~\cite{2018A&A...616A...1G})
allows the solving of several stellar astronomy problems,
like determination of
stellar mass, age estimation and others.
In~particular, it became possible to improve the results of the parameterization of stars,
carried out from multicolor~photometry.

In this paper, the verification of
the method and stellar sample analyzed in~\citet{2018OAst...27...62M}
using {\it Gaia} DR2 data is described. We also discuss how including the
{\it Gaia} parallaxes into the procedure
would improve the accuracy of parameterization, and how to select/process objects
with unknown {\it Gaia} parallax for parameterization.

This paper is organized as follows.
Data and methods are described in Section~\ref{sec:data},
Section~\ref{sec:results} contains results and conclusions.

\section{Data and Method}
\label{sec:data}

In~\citet{2018OAst...27...62M} (hereinafter---Paper18)
objects in four selected areas in the sky were cross-matched
(see details of the procedure in~\cite{2011ASPC..442..583M,2012AstBu..67...82K,2012BaltA..21..319M}) with
2MASS (Two Micron All-Sky Survey)~\cite{2006AJ....131.1163S},
SDSS (Sloan Digital Sky Survey)~\cite{2011ApJS..193...29A},
GALEX (Galaxy Evolution Explorer)~\cite{2005ApJ...619L...1M}, and
UKIDSS (UKIRT Infrared Deep Sky Survey)~\cite{2007MNRAS.379.1599L}
surveys, and multi-wavelength photometric data were used to determine the parameters of stars.
The galactic coordinates of the areas are
(334,  +61.9),
(257,  +48.7),
(301,  +62.1), and
(129,~$-$58.1).
For the studied objects MK (Morgan-Keenan)
spectral types (SpT), distances ($d$) and interstellar extinction values ($A_V$)
were estimated, minimizing the function
\begin{equation}
\chi^2 = \sum_{i=1}^N \left(\frac{m_{obs,i}-m_{calc,i}}{\sigma m_{obs,i}} \right)^2,
\label{equ:functional}
\end{equation}
where $m_{obs,i}$ and $\sigma m_{obs,i}$ are the apparent magnitude
and its observational error, respectively,
in the $i$-th photometric band from a~given survey, and
the summation is over up to {\it N} = 14 photometric bands
(JHK$_S$ from 2MASS, ugriz from SDSS, FUV and NUV from GALEX, YJHK from UKIDSS), and  
\begin{equation}
m_{calc,i} = M_i + 5\log{d} - 5 + A_i,
\label{equ:distance}
\end{equation}
or
\begin{equation}
m_{calc,i} = M_i - 5\log{\varpi} - 5 + A_i.
\label{equ:parallax}
\end{equation}

Here $A_i=f(A_V)$ is the extinction in the i-th photometric band,
and can be determined from the interstellar extinction law.
To retrieve $A_i$ from $A_V$
we have used data from \citet{2013MNRAS.430.2188Y} for
2MASS~\cite{2006AJ....131.1163S},
SDSS~\cite{2011ApJS..193...29A}
and GALEX~\cite{2005ApJ...619L...1M},
whereas data for UKIDSS~\cite{2007MNRAS.379.1599L} and Johnson V-band were adopted from
\citet{2011ApJ...737..103S}.
Both teams made $A_i$ calculations for SDSS photometry,
and our comparison shows a~very good agreement between their results
(see Paper18 for details).

$M_i=f(\rm{SpT})$ is the absolute magnitude in i-th photometric band taken
from calibration tables.
To obtain absolute
magnitudes for stars of different spectral types
in the corresponding photometric systems $M_i$,
we have compiled a~table of absolute magnitudes in 2MASS, SDSS and GALEX
surveys using
\citet{2007AJ....134.2340K}, and \citet{2011AJ....142...23F} data.
Due to lack of published data,
UKIDSS absolute magnitudes were calculated from 2MASS magnitudes
and 2MASS-UKIRT relations from \citet{2009MNRAS.394..675H}.

The distance $d$ and parallax $\varpi$
are expressed in parsec and arcsec, respectively.
It should be noted that Equation~\eqref{equ:parallax} is not correct when using observational data.
The mean value of the parallax is not enough and their errors
should be considered to derive a~good value for the distance
to be substituted in Equation~\eqref{equ:distance} to derive Equation~\eqref{equ:parallax}
(see~\citet{2018AJ....156...58B}).

Altogether 251 objects were found in at least three of the four surveys
and cross-matched in the four areas, but only 26 of them
were successfully parameterized. The following reasons to remove
objects from further consideration were considered.

First, the original surveys contain various flags which allow us
to remove unsuitable objects, namely,
``Binary object'' (2MASS, UKIDSS),
``Non-stellar/extended object'' (2MASS, SDSS, GALEX, UKIDSS),
``Observation of low quality'' (SDSS).
Secondly, overly bright objects and objects with
large observational errors were not considered.

Also, only
areas located at relatively high galactic latitudes ($|b|>48^{\circ}$)
are considered in this work.
Consequently, $A_V$ is assumed to be smaller than 0.5 mag, and
distance $d$ is assumed to be closer than 8000 pc. Objects which
presented larger values for those parameters were removed
from further~consideration.

Finally, for every object
a rough parameterization was performed (based on 2MASS+SDSS photometry only)
with \citet{2007AJ....134.2398C}
absolute magnitude tables. An object was removed if
this procedure showed that there was a~high probability for it
to be a~non-MS star (giant or supergiant). It~should be reminded
that in the current study
only MS (Main-Sequence) stars are considered.

A comparison of our results for 26 selected stars
with independent results obtained from the LAMOST
(Large Sky Area Multi-Object Fiber Spectroscopic Telescope)~\cite{2015RAA....15.1095L}
for some of the studied stars has demonstrated a~good agreement.
Interstellar extinction as a~function of distance ($A_V(d)$) was constructed
for the four selected areas (see Figures~1--4 in Paper18).
These relations were extrapolated to infinity ($d \to \infty$).
Please note that the extrapolation is formal, and can introduce some uncertainty.
For the resulting
$A_V(\infty)$ values
in three of the four areas (Nos 1, 5, and 6 in Paper18),
a good agreement was found with the data used in the study of
supernovae~\cite{1999ApJ...517..565P}
to confirm the accelerated expansion of the Universe.
For the remaining area, No 2, an agreement was not achieved
(see Table~2 in Paper18).

To identify the conditions under which
stars are properly parameterized from multicolor photometry,
we relied upon the results obtained
in our pilot study of interstellar extinction in four areas (see Paper18), and {\it Gaia} data.
A cross-matching of Paper18 objects with {\it Gaia} DR2 catalog was made.
Among 251 objects, studied in Paper18, only 72 were found in {\it Gaia} DR2
(such a~very small fraction is understandable, as about 80\%
of the Paper18 objects are fainter than $19.^m5$ $g_{SDSS}$),
and seven of them
have no {\it Gaia} parallaxes. Among 26 objects, selected in Paper18 for construction of
$A_V(d)$ relation, one is absent in {\it Gaia} DR2,
and two more objects have no parallaxes.

The {\it Gaia} trigonometric parallaxes (hereafter $\varpi_{tr}$)
are compared with photometric parallaxes obtained in Paper18
(hereafter $\varpi_{ph}$).
When analyzing {\it Gaia} data, the recommendations published
in~\cite{2018A&A...616A...2L} were considered. In particular, the filters,
designed on the basis of the photometric and astrometric flags contained in {\it Gaia} DR2,
were taken into account.
They were used to construct a~so-called astrometrically clean sample, hereafter ACS.

\begin{figure}[H]
\centering
\includegraphics[width=10 cm]{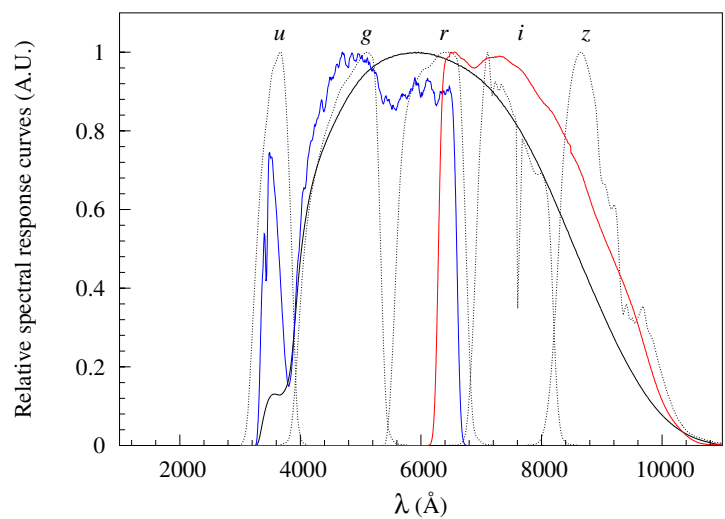}
\caption{SDSS ugriz (gray dashed curves) and {\it Gaia} G (black solid curve),
G$_{BP}$ (blue solid curve), and~G$_{RP}$ (red solid curve) photometric passbands.}
\label{fig:bands}
\end{figure}

Photometric passbands used in SDSS~\cite{2010AJ....139.1628D}
and {\it Gaia} DR2~\cite{2018A&A...616A...4E} are presented in Figure~\ref{fig:bands}.
In this work, we deal with g$_{SDSS}$ and {\it Gaia} G, BP, and RP magnitudes
of the studied stars.

\section{Results and Conclusions}
\label{sec:results}

\begin{figure}[H]
\centering
\includegraphics[width=8.5 cm]{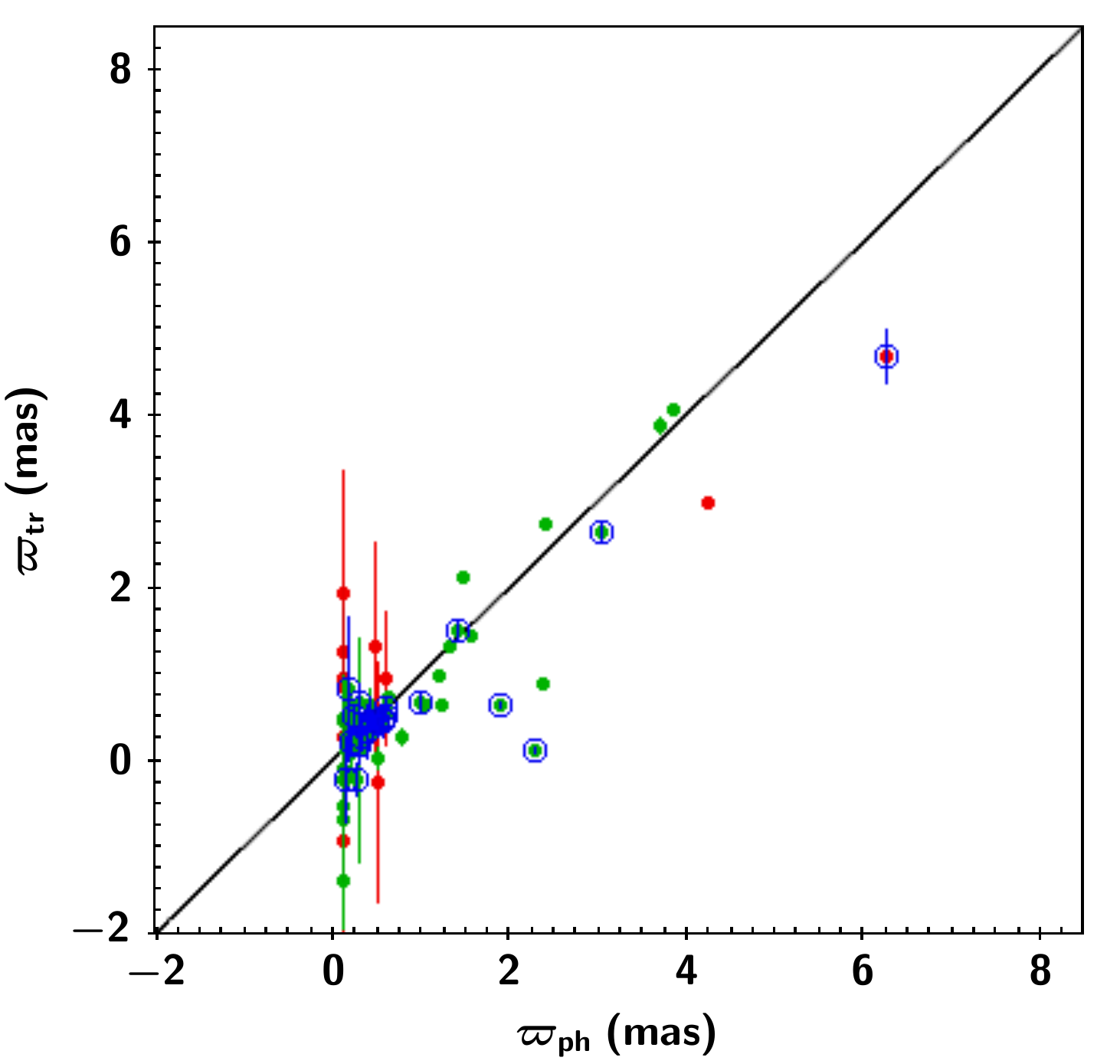}
\caption{
Photometric parallax ($\varpi_{ph}$) from Paper18 and trigonometric parallax
($\varpi_{tr}$) from {\it Gaia} for
all stars in common to Paper18 and {\it Gaia} DR2 that satisfy (green points)
and do not satisfy (red points) the ACS requirements.
Blue circles are the stars used in Paper18 for the $A_V(d)$ construction.
``$\varpi_{tr}$ = $\varpi_{ph}$'' is indicated as a~black line.
}
\label{fig:plx}
\end{figure}

\begin{figure}[H]
\centering
\includegraphics[width=8.5 cm]{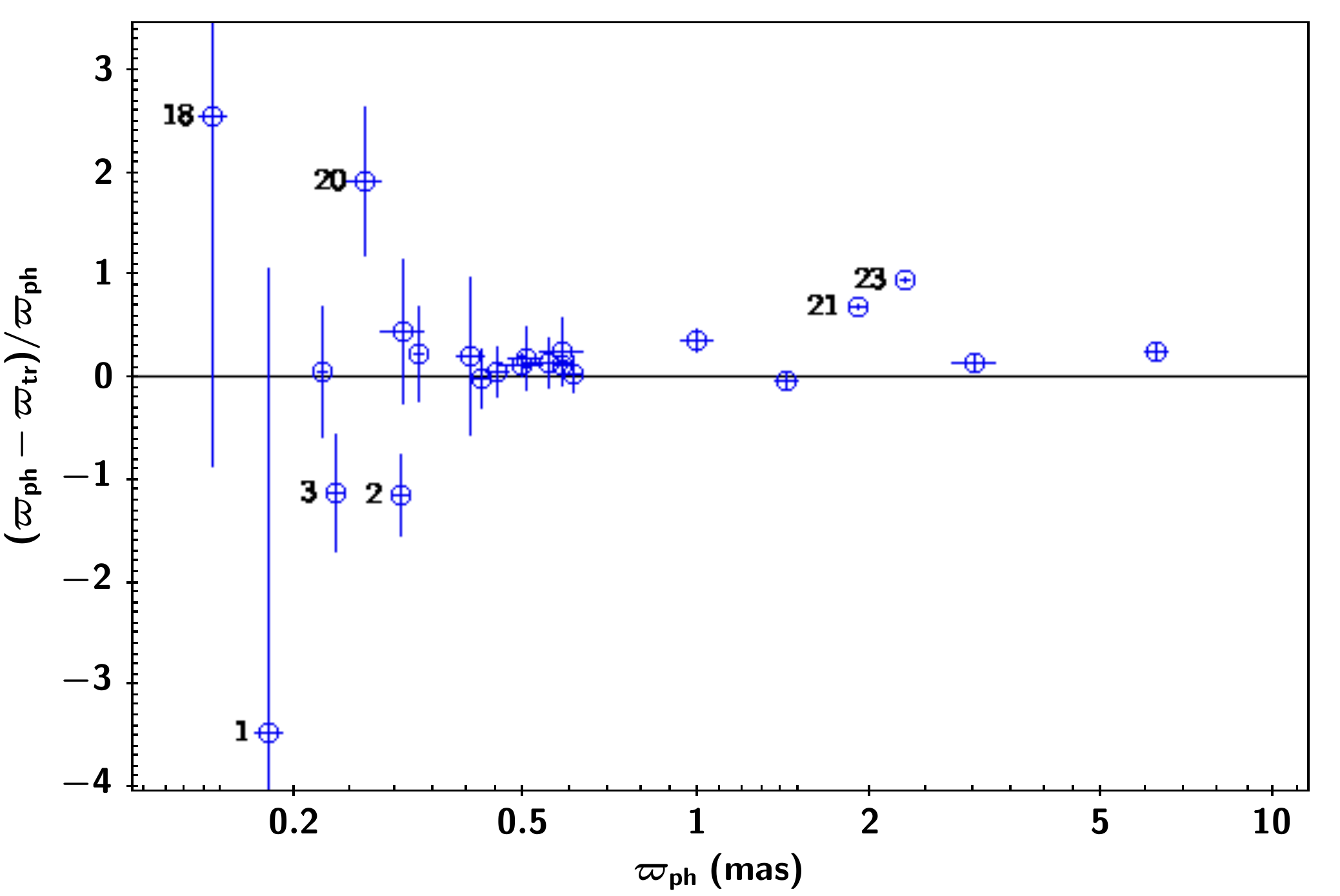}
\includegraphics[width=8.5 cm]{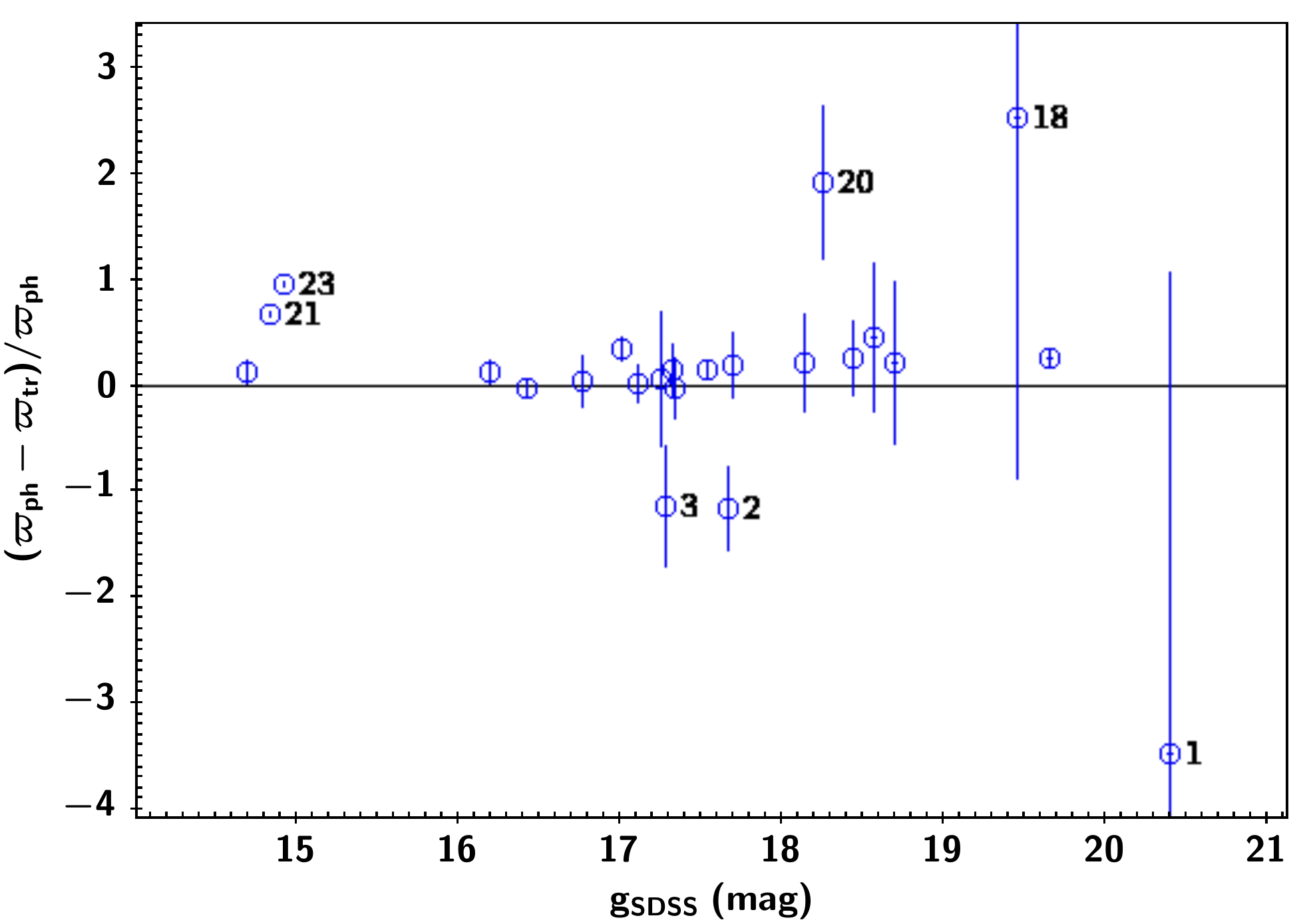}
\caption{
The ratio between the difference in parallax $\varpi_{ph} - \varpi_{tr}$
and the photometric parallax $\varpi_{ph}$ vs. $\varpi_{ph}$
(\textbf{upper} panel) and vs. $g_{SDSS}$ (\textbf{lower} panel)
for the stars used in Paper18
for the $A_V(d)$ construction. Labels correspond to running numbers of stars
in Table~\ref{tab:stars}.}
\label{fig:plxratio}
\end{figure}

A comparison of the photometric and trigonometric parallaxes of stars
used in Paper18 for the $A_V(d)$ construction is shown in Figure~\ref{fig:plx}.
The ratio between the difference in parallax $\varpi_{ph} - \varpi_{tr}$
and the photometric parallax $\varpi_{ph}$
as a~function of $\varpi_{ph}$ and $g_{SDSS}$
is shown in Figure~\ref{fig:plxratio}.
In the current study observational photometric errors are considered to be
the only source for the resulting parameters errors.
Consequently, here we underestimate the error values. To calculate errors more correctly,
one~should also consider calibration tables errors and relations errors.

\begin{figure}[H]
\centering
\includegraphics[width=10 cm]{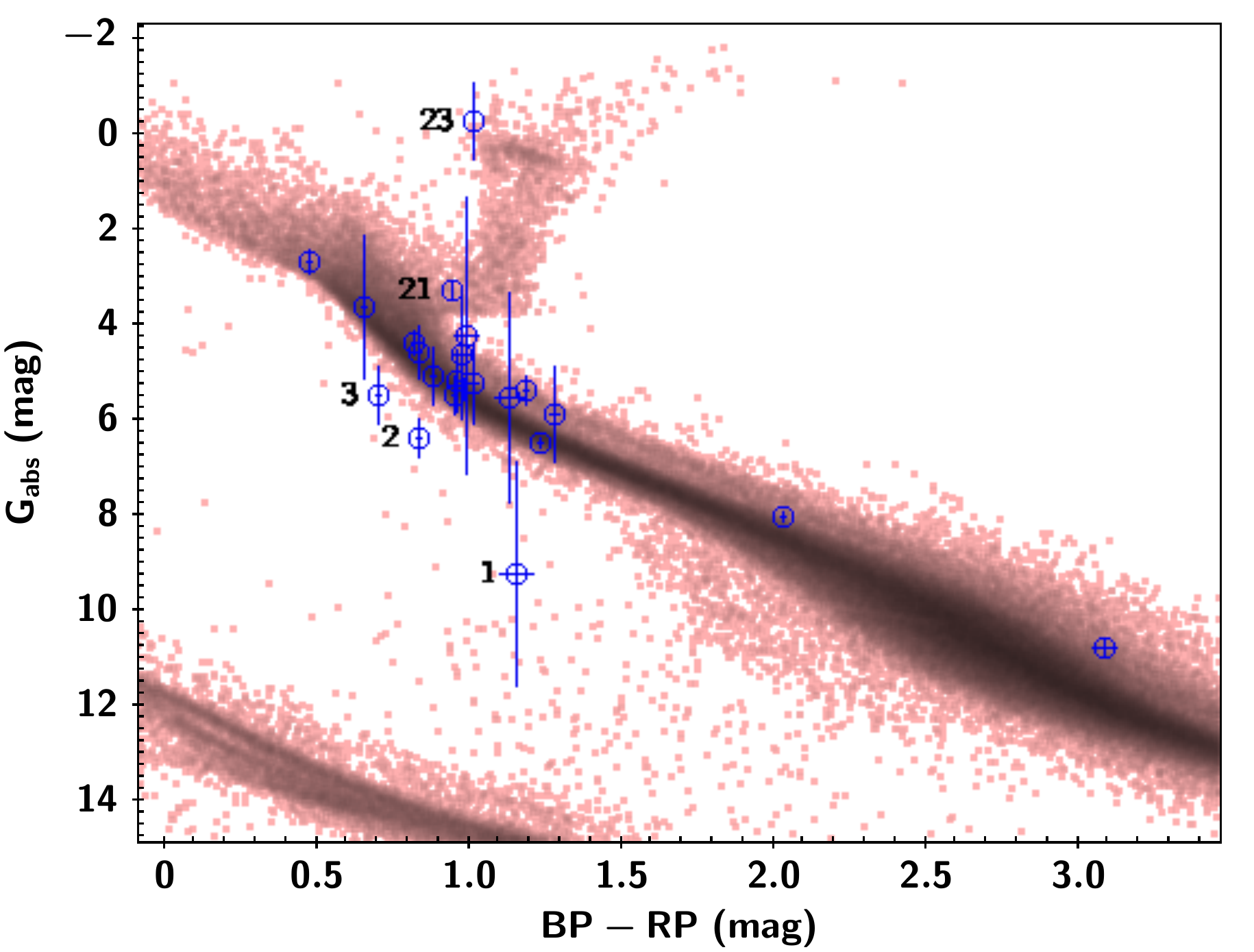}
\caption{Stars used in Paper18 for the $A_V(d)$ construction
(blue circles) on the HRD.
Labels correspond to running numbers of stars
in Table~\ref{tab:stars}.
Pink points are nearest ($\varpi_{tr} > 10$ mas) stars from {\it Gaia} DR2.
Absolute magnitude $G_{abs}$ is calculated from $G_{Gaia}$ and $\varpi_{tr}$,
interstellar extinction is neglected. {\it Gaia} G, BP and RP curves are presented
in Figure~\ref{fig:bands}.}
\label{fig:hrd}
\end{figure}

Most stars demonstrate a~satisfactory agreement but with some
outliers which require an explanation. The Hertzsprung-Russell diagram
(HRD) was constructed for {\it Gaia} DR2 stars
with trigonometric parallax $\varpi_{tr} > 10$ mas,
with relative parallax uncertainty $\sigma\varpi_{tr} /\varpi_{tr} < 10\%$,
with relative error of BP and RP fluxes better than 10\%,
and satisfying the ACS requirements (Figure~\ref{fig:hrd}).
The studied stars were added to the plot, and their
positions on HRD were analyzed.
Data for the studied stars are presented in Table~\ref{tab:stars}.
\unskip

\begin{table}[H]
\caption{Stars used in Paper18 for the $A_V(d)$ construction.
Upper part: running number, SDSS and {\it Gaia} identifiers, RA, and DEC (2000)
coordinates. Lower part: running number, SDSS ($g$)
and {\it Gaia} ($G$) photometry with
errors, photometric and trigonometric parallax with errors,
spectral type estimated in Paper18 and corresponding $M_{bol}$.
}
\centering
\scalebox{1}[1]{\begin{tabular}{ccccccccccc}
\toprule
\multicolumn{1}{c}{\textbf{\emph{N}}} &
\multicolumn{3}{c}{\textbf{SDSS}} &
\multicolumn{3}{c}{\textbf{{\emph{Gaia}} DR2}} &
\multicolumn{2}{c}{\textbf{RA2000 (deg)}} &
\multicolumn{2}{c}{\textbf{DEC2000 (deg)}} \\
\midrule
1 &  \multicolumn{3}{c}{752-301-2-0325-0286} & \multicolumn{3}{c}{3683644191076349312} & \multicolumn{2}{c}{192.1701}&\multicolumn{2}{c}{ $-$0.8165} \\
2 & \multicolumn{3}{c}{6005-301-2-0158-0092} &\multicolumn{3}{c}{ 3689648868888993920 }& \multicolumn{2}{c}{192.2168}&\multicolumn{2}{c}{ $-$0.7911} \\
3 & \multicolumn{3}{c}{6005-301-2-0158-0061} &\multicolumn{3}{c}{ 3683643881838700800} & \multicolumn{2}{c}{192.1728}& \multicolumn{2}{c}{$-$0.8335} \\
4 & \multicolumn{3}{c}{1241-301-1-0173-0042} & \multicolumn{3}{c}{3801228953847961600} & \multicolumn{2}{c}{164.2724}& \multicolumn{2}{c}{$-$3.5562} \\
5 &  \multicolumn{3}{c}{752-301-2-0324-0052 }&\multicolumn{3}{c}{ 3683656594941905920 }& \multicolumn{2}{c}{192.0763}& \multicolumn{2}{c}{$-$0.7852} \\
6 & \multicolumn{3}{c}{1462-301-4-0543-0082} & \multicolumn{3}{c}{3665039324758041600} & \multicolumn{2}{c}{206.7882}& \multicolumn{2}{c}{ 2.4288 }\\
7 & \multicolumn{3}{c}{6121-301-2-0157-0087 }& \multicolumn{3}{c}{3683657110337986048 }& \multicolumn{2}{c}{192.1002}&\multicolumn{2}{c}{ $-$0.7568 }\\
8 & \multicolumn{3}{c}{6121-301-2-0157-0037} & \multicolumn{3}{c}{3683644569033472512 }& \multicolumn{2}{c}{192.1050}&\multicolumn{2}{c}{ $-$0.8108 }\\
9 & \multicolumn{3}{c}{1462-301-4-0543-0066 }& \multicolumn{3}{c}{3665227032008658560} & \multicolumn{2}{c}{206.7471}&  \multicolumn{2}{c}{2.4403} \\
10 & \multicolumn{3}{c}{2194-301-1-0361-0007} &\multicolumn{3}{c}{ 3801228683265104640} & \multicolumn{2}{c}{164.1956}&\multicolumn{2}{c}{ $-$3.5772} \\
11 & \multicolumn{3}{c}{1241-301-1-0173-0102} & \multicolumn{3}{c}{3801228610250558848 }&\multicolumn{2}{c}{ 164.2084}&\multicolumn{2}{c}{ $-$3.5842} \\
12 & \multicolumn{3}{c}{7727-301-3-0174-0108 }& \multicolumn{3}{c}{2552008097711479808 }& \multicolumn{2}{c}{ 16.2328}&  \multicolumn{2}{c}{4.5274} \\
13 &\multicolumn{3}{c}{ 2194-301-1-0361-0119} & \multicolumn{3}{c}{3801227819976566272} & \multicolumn{2}{c}{164.2968}& \multicolumn{2}{c}{$-$3.5978} \\
14 &  \multicolumn{3}{c}{752-301-2-0325-0136} &\multicolumn{3}{c}{ 3689650277638267392 }&\multicolumn{2}{c}{ 192.1738}& \multicolumn{2}{c}{$-$0.7430} \\
15 & \multicolumn{3}{c}{1462-301-4-0543-0182} & \multicolumn{3}{c}{3665226581036533504 }& \multicolumn{2}{c}{206.7508}&\multicolumn{2}{c}{  2.4200} \\
16 &  \multicolumn{3}{c}{756-301-1-0510-0121} & \multicolumn{3}{c}{3683643679975511808 }& \multicolumn{2}{c}{192.1394}& \multicolumn{2}{c}{$-$0.8415 }\\
17 & \multicolumn{3}{c}{7727-301-3-0174-0122} &\multicolumn{3}{c}{ 2552021880262020608} &  \multicolumn{2}{c}{16.2512}&  \multicolumn{2}{c}{4.6030} \\
18 & \multicolumn{3}{c}{1462-301-4-0544-0120} & \multicolumn{3}{c}{3665040626132812800} &\multicolumn{2}{c}{ 206.8532}& \multicolumn{2}{c}{ 2.4438} \\
19 &\multicolumn{3}{c}{ 7727-301-3-0175-0149} & \multicolumn{3}{c}{2552009442036732416} &  \multicolumn{2}{c}{16.2885}&  \multicolumn{2}{c}{4.5344 }\\
20 &  \multicolumn{3}{c}{752-301-2-0324-0183} & \multicolumn{3}{c}{3689662230532099200} & \multicolumn{2}{c}{192.0995}&\multicolumn{2}{c}{ $-$0.7200} \\
21 & \multicolumn{3}{c}{1462-301-4-0543-0079} &\multicolumn{3}{c}{ 3665039290398303104} &\multicolumn{2}{c}{ 206.7741}& \multicolumn{2}{c}{ 2.4244 }\\
22 & \multicolumn{3}{c}{2194-301-1-0361-0077} & \multicolumn{3}{c}{3801228644610295680} & \multicolumn{2}{c}{164.1954}& \multicolumn{2}{c}{$-$3.5860 }\\
23 & \multicolumn{3}{c}{1462-301-4-0543-0060} & \multicolumn{3}{c}{3665226615396253824} &\multicolumn{2}{c}{ 206.7224}& \multicolumn{2}{c}{ 2.4265} \\
\midrule
%
%
\multicolumn{1}{c}{{\textbf{N}}} &
\multicolumn{1}{c}{\boldmath{$g$}} &
\multicolumn{1}{c}{\boldmath{$\sigma_g$}} &
\multicolumn{1}{c}{\boldmath{$G$}} &
\multicolumn{1}{c}{\boldmath{$\sigma_G$}} &
\multicolumn{1}{c}{\boldmath{$\varpi_{ph}$}} &
\multicolumn{1}{c}{\boldmath{$\sigma \varpi_{ph}$}} &
\multicolumn{1}{c}{\boldmath{$\varpi_{tr}$}} &
\multicolumn{1}{c}{\boldmath{$\sigma \varpi_{tr}$}} &
\multicolumn{1}{c}{\textbf{Sp}} &
\multicolumn{1}{c}{\boldmath{$M_{bol}$}} \\
\midrule
1 & 20.384 & 0.021 & 19.718 & 4.3 $\times$ 10$^{-3}$ & 0.1811 & 0.0097 &  0.8123 & 0.821  & G8 & 5.3 \\ 
2 & 17.674 & 0.005 & 17.260 & 8.4 $\times$ 10$^{-4}$ & 0.3086 & 0.0115 &  0.667  & 0.1211 & G0 & 4.47\\
3 & 17.269 & 0.005 & 16.974 & 8.7 $\times$ 10$^{-4}$ & 0.2375 & 0.0067 &  0.5096 & 0.1363 & F5 & 3.61\\
4 & 16.408 & 0.003 & 15.646 & 1.4 $\times$ 10$^{-3}$ & 1.4285 & 0.0592 &  1.4892 & 0.0784 & K2 & 6.08\\
5 & 17.335 & 0.005 & 16.882 & 7.4 $\times$ 10$^{-4}$ & 0.4255 & 0.0168 &  0.4383 & 0.1179 & G5 & 4.89\\
6 & 17.076 & 0.005 & 16.597 & 1.2 $\times$ 10$^{-3}$ & 0.6116 & 0.0219 &  0.6022 & 0.1014 & G8 & 5.3 \\
7 & 17.251 & 0.005 & 16.998 & 7.8 $\times$ 10$^{-4}$ & 0.2252 & 0.0047 &  0.2149 & 0.1413 & F5 & 3.61\\
8 & 16.788 & 0.004 & 16.387 & 6.3 $\times$ 10$^{-4}$ & 0.4524 & 0.0191 &  0.4361 & 0.1048 & G0 & 4.47\\
9 & 14.645 & 0.003 & 14.512 & 6.5 $\times$ 10$^{-4}$ & 0.4975 & 0.0421 &  0.439  & 0.0439 & F2 & 2.89\\
10 & 16.200 & 0.003 & 15.816 & 7.9 $\times$ 10$^{-4}$ & 0.5882 & 0.0159 &  0.5176 & 0.0594 & G0 & 4.47\\
11 & 18.110 & 0.006 & 17.596 & 1.9 {$\times$ 10$^{-3}$} & 0.3300 & 0.0066 &  0.2585 & 0.1536 & G5 & 4.89\\
12 & 17.327 & 0.005 & 16.824 & 1.0 $\times$ 10$^{-3}$ & 0.5571 & 0.0541 &  0.4783 & 0.1317 & G8 & 5.3 \\
13 & 18.699 & 0.009 & 18.025 & 3.2 {$\times$ 10$^{-3}$} & 0.4048 & 0.0206 &  0.3235 & 0.3079 & K0 & 5.69\\
14 & 17.689 & 0.005 & 17.169 & 8.9 $\times$ 10$^{-4}$ & 0.5102 & 0.0343 &  0.4149 & 0.153  & G8 & 5.3 \\
15 & 18.616 & 0.008 & 18.106 & 2.0  {$\times$ 10$^{-3}$} & 0.3095 & 0.0258 &  0.1714 & 0.217  & G8 & 5.3 \\
16 & 18.440 & 0.007 & 17.658 & 1.2  {$\times$ 10$^{-3}$} & 0.5865 & 0.0510 &  0.4417 & 0.1949 & K2 & 6.08\\
17 & 17.004 & 0.004 & 16.281 & 9.1 $\times$ 10$^{-4}$ & 1.0050 & 0.0601 &  0.6621 & 0.0933 & K2 & 6.08\\
18 & 19.437 & 0.012 & 19.017 & 4.1  {$\times$ 10$^{-3}$} & 0.1455 & 0.0075 & $-$0.2226 & 0.4955 & G0 & 4.47\\
19 & 17.547 & 0.005 & 15.965 & 9.4 $\times$ 10$^{-4}$ & 3.0303 & 0.2470 &  2.6283 & 0.102  & M0 & 7.6 \\
20 & 18.248 & 0.006 & 17.784 & 1.2  {$\times$ 10$^{-3}$} & 0.2673 & 0.0174 & $-$0.2431 & 0.1933 & G0 & 4.47\\
21 & 14.820 & 0.003 & 14.332 & 4.9 $\times$ 10$^{-4}$ & 1.9047 & 0.0447 &  0.626  & 0.0398 & G8 & 5.3 \\
22 & 19.643 & 0.015 & 17.431 & 7.7  {$\times$ 10$^{-3}$} & 6.25   & 0.2734 &  4.6726 & 0.2955 & M4 & 9.92\\
23 & 14.901 & 0.003 & 14.361 & 3.7 $\times$ 10$^{-4}$ & 2.2988 & 0.0497 &  0.1203 & 0.0418 & K2 & 6.08\\
\bottomrule
\end{tabular}}

\label{tab:stars}
\end{table}

Stars 21 and 23 belong to red giants. Obviously, they were wrongly accepted as
Main-Sequence stars in Paper18, which resulted in erroneous $\varpi_{ph}$ values.

According to Figure~\ref{fig:hrd}, stars 1, 2 and 3 are sub-dwarfs,
and, if that is true, they have erroneous $\varpi_{ph}$ for the same reason
as above.
However, we pay attention to their relatively large trigonometric parallax errors
(see $\sigma \varpi_{tr}$ values in Table~\ref{tab:stars}), which could lead to
their shift under the Main Sequence in~HRD.

Lastly, stars 18 and 20 have negative parallaxes and for that reason their
absolute magnitudes cannot be calculated and, consequently, they are not shown
in Figure~\ref{fig:hrd}. Apparently the negative parallaxes indicate that
these stars belong to supergiants. Again, they were wrongly accepted to be
Main-Sequence stars in Paper18, which resulted in erroneous $\varpi_{ph}$ values.

The results of the comparison allow us to make the following conclusions.
\begin{enumerate}[leftmargin=*,labelsep=5mm]
\item A parameterization of stars with
$\varpi_{ph}~>~0.225$ mas
(i.e., closer than about 4400 pc)
and $g_{SDSS}~<~19.^m6$
(see Figure~\ref{fig:plxratio}) is successful, subject to a~proper determination
of luminosity~class.
\item
A rough estimate of the probability of stars belonging to the Main Sequence
was carried out in Paper18 (basing on 2MASS and SDSS photometry data),
and only MS stars were parameterized.
Obviously, for several stars that estimation turned out to be erroneous
(see Figure~\ref{fig:hrd}).
It seems appropriate to include in the parameterization procedure
information about the photometry of non-MS stars (sub-dwarfs, giants and supergiants),
drawn from the literature or determined by our own efforts.
\end{enumerate}

\begin{figure}[H]
\centering
\includegraphics[width=10 cm]{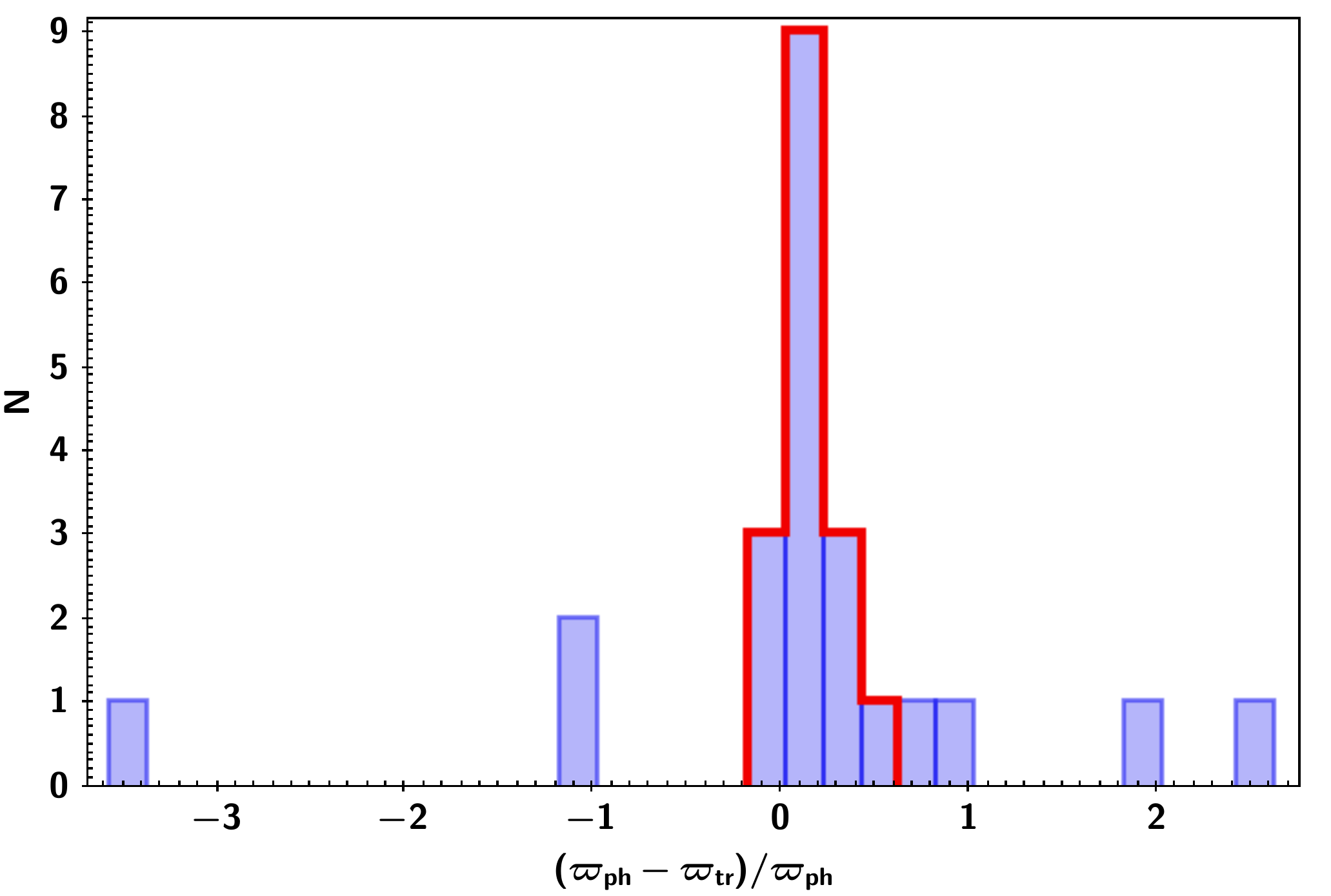}
\caption{Distribution of
stars used in Paper18 for the $A_V(d)$ construction
by the ratio between the difference in parallax $\varpi_{ph} - \varpi_{tr}$
and the photometric parallax $\varpi_{ph}$. Subsample, designated
by the red contour, does not include
stars 21, 23 (red giants), 1, 2, 2 (possible sub-dwarfs)
and 18, 20 (apparently supergiants),
discussed in the text.}
\label{fig:histogram}
\end{figure}

The distribution of
stars used in Paper18 for the $A_V(d)$ construction
by the ratio between the difference in parallax $\varpi_{ph} - \varpi_{tr}$
and the photometric parallax $\varpi_{ph}$ is presented in Figure~\ref{fig:histogram}.
Seven outliers demonstrate $|(\varpi_{ph} - \varpi_{tr})/\varpi_{ph}| > 0.6$,
they were discussed above (stars 1, 2, 3, 18, 20, 21, 23).

The remaining stars distribution is shown with the red contour.
For them mean value of $(\varpi_{ph} - \varpi_{tr})/\varpi_{ph}$
is 0.15 with standard deviation of 0.13. Mean $\varpi_{ph}$ and
$\varpi_{tr}$ values for the red contour group are 0.997 and 0.89,
respectively.

It should be noted that area No 2, which showed a~poor agreement with the
Perlmutter et al.'s~\cite{1999ApJ...517..565P} data in Paper18,
does not stand out in the current analysis.

It should be also noted
that the cross-matching of Paper18 objects with {\it Gaia} DR2 catalog was made correctly.
Angular distance on the sky between a~Paper18 object and
its {\it Gaia} DR2 counterpart ($\rho$) does not exceed 1 arcsec, and only
for 2 of 72 objects $\rho>0.3$ arcsec (those two stars were not selected in Paper18 for
construction of $A_V(d)$ relation, and consequently, are not shown
in Figures~\ref{fig:plxratio}--\ref{fig:histogram}).

Based on the conclusions derived above,
our procedure of parameterization of stars
will be modified. In particular,
the procedure will be extended to sub-dwarf, giant and supergiant stars.
Also, at this stage,
distant ($\varpi_{ph} < 0.225$ mas) and faint ($g_{SDSS} > 19.^m6$) objects
will be removed from consideration.
We will also reconstruct the procedure to use
{\it Gaia DR2} parallaxes (and future releases, when available),
as an input parameter.
It is also planned to use data from other multicolor surveys,
such as~WISE (Wide-Field Infrared Survey Explorer),
DENIS (Deep Near Infrared Survey of the Southern Sky), etc.,
and extend this procedure to lower
galactic latitudes.

\vspace{6pt}

\authorcontributions{The authors made equal contribution to this work; O.M. wrote the paper.}

\funding {This research was partly funded by
the Russian Foundation for Basic Research grant 17-52-45076.
A.K.~acknowledges the National Research Foundation of South Africa and
the Russian Science Foundation (project no. 14-50-00043).
G.Z. acknowledges support by NSFC grant No. 11390371.}

\acknowledgments{
We are grateful to our reviewers whose constructive comments
greatly helped us to improve the paper.
We thank E. Kilpio for fruitful collaboration.
This work has made use of data from the European Space Agency (ESA) mission
{\it Gaia} (\url{https://www.cosmos.esa.int/gaia}), processed by the {\it Gaia}
Data Processing and Analysis Consortium (DPAC,
\url{https://www.cosmos.esa.int/web/gaia/dpac/consortium}). Funding for the DPAC
has been provided by national institutions, in particular the institutions
participating in the {\it Gaia} Multilateral~Agreement.
}

\conflictsofinterest{The authors declare no conflict of interest.}




\reftitle{References}




\end{document}